\documentstyle[aps,multicol,citesort,psfig]{revtex}

\newcommand{\ageq}{\mbox{\
\raisebox{-.9ex}{$\stackrel{\textstyle >}{\sim}$}\ }}

\begin{document}
\bibliographystyle{prsty}

\title{Twirling Elastica:  Kinks, Viscous Drag, and Torsional Stress}

\author{Stephan A. Koehler$^1$ and Thomas R. Powers$^2$}
\address{$^1$Division of Engineering and Applied Sciences, Harvard University, 
Cambridge, MA  02138}
\address{$^2$Division of Engineering, Brown University, Providence, RI 02912}

\date{6 April 2000; revised 28 September 2000}
  
\maketitle

\begin{abstract}
\noindent
Biological filaments such as DNA or bacterial flagella are typically
curved in their natural states.  To elucidate the interplay of 
viscous drag, twisting, and bending in the overdamped dynamics
of such filaments, we 
compute the steady-state 
torsional stress and shape of a rotating rod with a kink.  Drag 
deforms the rod, ultimately extending or folding it depending on 
the kink angle.  For certain kink angles and kink locations,
both states are possible at high rotation rates.  The agreement between our 
macroscopic experiments and the theory is good, with no adjustable parameters.

\end{abstract}




\begin{multicols}{2}
The coupling of viscous stresses from fluid flow to deformations of elastic
fibers is important in many situations, such paper manufacture, flexible microstructures in MEMS devices, and the
dynamics of flexible biological filaments.  
The last example includes spinning filiments, such as
bacterial flagella, DNA, and supercoiling
colonies of mutant strains of {\it Bacillus subtilis}.  
The rotary motors of {\it E. coli} rotate the 
flagellar bundle up to $9000$ rpm~\cite{alberts}.  
In DNA transcription, if rotational motion of the 
RNA polymerase is blocked, then 
the DNA twirls at typical rates of $300$--$600$ rpm as it is 
pulled through the enzyme~\cite{ikeda}.  
Finally, the mutant {\it B. subtilis} cells fail to separate as they divide,
and form long fibers which rotate as they grow~\cite{nm}.

Although these filaments are often modeled as intrinsically straight 
elastic rods, natural bends can have significant effects.
For example, natural curvature in bacterial flagella is crucial for generating thrust; mutants with straight flagella can't swim~\cite{macnab}.
Proteins can bind to DNA and 
impose sharp, large-angle bends~\cite{kim}, which can greatly 
enhance the torsional stress due to viscous drag during transcription.
Similar torsional stresses from intrinsic bends have been estimated to
be large enough to affect gene activity or DNA structure even in the 
absence of external anchoring~\cite{pcn}.  As we shall recall below,
kinks generically {\it trap} torsional stress 
in specific regions of a rotating filament.    Finally, there 
is evidence that intrinsic bends may arise at the hairpin loops 
during the formation of supercoiled colonies of {\it B. subtilis}~\cite{nm1}.

To gain intuition on how sharp intrinsic bends affect shape and twist
in the overdamped (inertialess) regime of cellular motions,
we study a bent elastic rod rotating in an extremely viscous fluid,
and ask, what is the shape and torsional stress as a function of twirling
rate?  For simplicity we disregard Brownian effects.
We first describe how rotation affects the shape, then we formulate and 
solve the problem using slender-body theory, and finally we compare
our predictions with experimental results.

The shape and stresses of a rotating elastic rod 
in a viscous fluid depend on 
its stress-free state.  At low rotation rates, a naturally
straight rod twirled about its long axis (which is along the $z$-axis, say) 
remains straight, but twists and spins about $z$.
At higher rates this state is unstable, the centerline writhes and 
slowly rotates about $z$, and each element of
the rod rapidly spins about the local tangent.  This motion is a hybrid
of crankshaft and speedometer-cable motion~\cite{wpg}.
\begin{figure}
\centerline{\psfig{figure=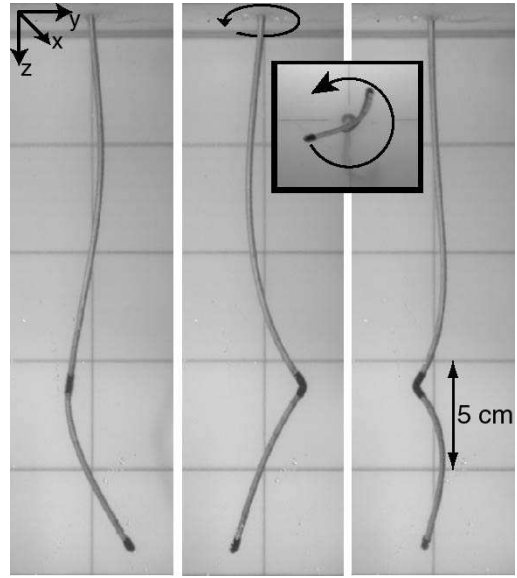,height=3.0in}}
\smallskip
\caption{Time sequence, from left to right, showing the steady-state shape
of a naturally L-shaped rod twirling in glycerol.  
A motor (above, not shown) rotates the rod at 200 rpm.
Gravity is along $z$, and $x$ points into the page.
The container is a Plexiglas box 
$31$ cm $\times 31$ cm wide and 28 cm tall.  Inset is view from below, in
which the rod rotation is counter-clockwise.}
\label{L}
\end{figure}

\begin{figure}
\centerline{\psfig{figure=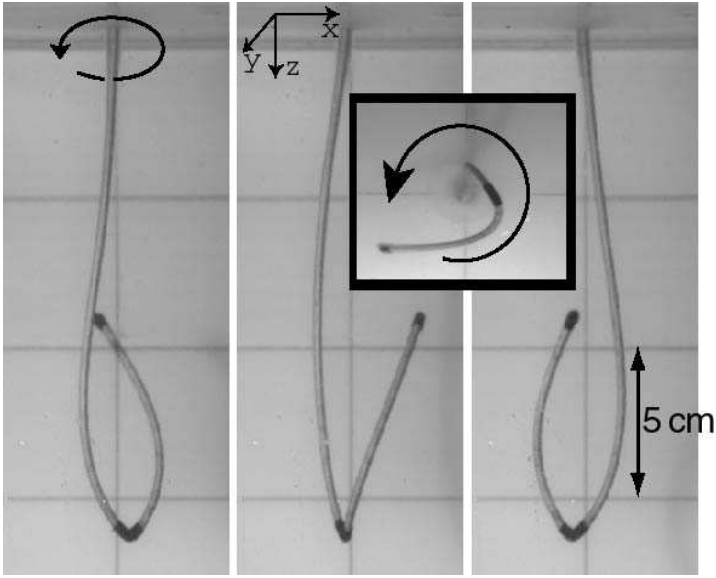,height=3.0in}}
\smallskip
\caption{Bistability at sufficiently high $\omega$.
The same rod as in Fig. 1, spinning at $200$ rpm,
but a different state, obtained by bending the rod upwards into a
V-shape at the onset of twirling.  Inset:  view from below.
}
\label{V}
\end{figure}

A naturally bent rod rotating in a viscous fluid behaves very differently.
Consider a rod made of two straight legs joined at a right angle.
Align one leg along the vertical ($z$-) axis 
and twirl it with velocity $\omega{\bf\hat z}$.
Unlike the naturally straight rod, the centerline of this
rod will distort from its unstressed state for {\it any} rotation rate, 
since the free leg
experiences translational drag.  This translational drag will wrap the
free leg around $z$ (insets, Fig.~\ref{L}--\ref{V}), and
twist the held leg.  In comparison to the naturally straight rod, the
torsional stress in the held leg is very large, for a given $\omega$.

Viscous stresses tend to straighten out a rod with a right-angle 
kink (Fig.~\ref{L}).
To see why, it is convenient to work in
the frame of the twirling rod, with an unperturbed fluid velocity
${\bf v}=\omega{\bf \hat z}\times{\bf r}$, where
{\bf r}=${\bf r}(s,t)$ is the position of the rod at time $t$, and
$s$ is arclength.
Suppose the free leg aligns parallel to the $x$-axis 
for $\omega=0$; {\it i.e.}, the ``L'' lies in the $z$-$x$ plane.
For small $\omega$ (``small'' will be defined below), 
translational drag deflects the free leg in 
the $y$-direction, which 
causes the held leg to bend away from the rotation axis in the $y$-direction.
Thus the held leg lies in a region where the flow is in the $-x$-direction;
this flow deflects the held leg, rotating the joint to extend the rod.
For small $\omega$, the sense of the joint rotation is {\it independent} 
of the exterior kink angle, $\alpha$.   
As $\omega$ increases,
rods with small to moderate $\alpha$ (L-shaped) 
will extend, whereas rods with large $\alpha$ (V-shaped)
will fold.   

These arguments suggest there should be a critical angle dividing the 
ultimate (large $\omega$) extending
and folding behaviors, but further reflection reveals that both states
are possible at large rotation rates.  Imagine that a rod with a right-angle 
kink is bent by an external force that brings the two legs together
so the rod looks like a ``U'' with a sharp kink.
Sufficient rotational flow will cause folding for the
same reason the V-shaped rod eventually folds.  
At high rotation rates, the 
viscous stresses will be large enough to hold the rod in this folded
state even when the external force is released.  Thus, we expect bistability
at large rotation rates: Figs.~\ref{L} and \ref{V} show the same rod and same
rotation rate with different initial conditions.
These steady-state 
shapes are stable against small-amplitude perturbations, but sufficiently 
large perturbations lead to a transition from extended to folded and 
vice-versa.

We now turn to the mathematical analysis.  For a slender rod 
(radius $a$ much less than length $L$), deformations can be
described by the configuration of an orthonormal material frame 
$\{{\bf \hat e}_1,{\bf\hat e}_2, {\bf\hat e}_3\}$ embedded in the 
rod~\cite{love}, with 
${\bf\hat e}_3=\partial{\bf r}/\partial s\equiv{\bf r}_s$.  
The rates of rotation
of this frame with respect to arclength and time define the angular strain
${\mbox{\boldmath$\Omega$}}$ and velocity ${\mbox{\boldmath$\omega$}}$:
\begin{equation}
({\bf\hat e}_i)_s= {\mbox{\boldmath$\Omega$}}\times{\bf\hat e}_i,
\quad\quad\quad
\label {omegas}
({\bf\hat e}_i)_t= {\mbox{\boldmath$\omega$}}\times{\bf\hat e}_i,
\label{omegat}
\end{equation}
with $i=1$--$3$.
Associated with these strains are internal elastic stresses, which
give rise to a force ${\bf F}(s)$ and a moment ${\bf M}(s)$ acting on 
the cross section at $s$. 
For an isotropic linearly elastic rod, the constitutive relation is 
${\bf M}=A{\bf r}_s\times{\bf r}_{ss}
+C{\bf r}_s\ {\bf r}_s\cdot{\mbox{\boldmath$\Omega$}}$, 
where $A$ is the bending
stiffness and $C$ is the twisting stiffness~\cite{love}.  Note 
${\bf M}=A{\mbox{\boldmath$\Omega$}}$ when $\Gamma\equiv C/A=1$.
In the (overdamped) limit of zero Reynolds number $Re$, 
the elastic force and 
moment per unit length balance exactly with the viscous
drag force and moment per unit length:
\begin{eqnarray}
{\bf F}_s-{\bf f}&=&{\bf 0},\\
\label{fbalance}
{\bf M}_s+{\bf r}_s\times{\bf F}-{\bf m}&=&{\bf 0}.
\label{mbalance}
\end{eqnarray}
From slender-body hydrodynamics~\cite{kr}, the leading-order drag force 
and moment per unit length are
\begin{eqnarray}
{\bf f}&=&\zeta_{\parallel}{\bf r}_s\ {\bf r}_s\cdot{\bf r}_t+\zeta_{\perp}(
{\bf r}_t-{\bf r}_s\ {\bf r}_s\cdot{\bf r}_t),\\
{\bf m}&=&\zeta_r{\bf r}_s\ {\bf r}_s\cdot{\mbox{\boldmath$\omega$}},
\label{drag}
\end{eqnarray}
where $\zeta_\perp\approx2\zeta_\parallel\approx4\pi\eta/[\log(L/2a)+c]$
($\eta$ is viscosity, $c$ is a constant of order unity) and 
$\zeta_r\approx4\pi\eta a^2$~\cite{kr}.  Although the slender-body 
approximation is invalid near the kink where the curvature
diverges, it does have the correct (linear) $\omega$-dependence for
the translational drag per unit length, and will 
therefore give the correct (nonlinear) scaling of the shape 
and torsional stress with $\omega$.  

As emphasized in ref.~\cite{pcn}, kinks block speedometer-cable
motion, since sharp curvature implies a large elastic energy
cost for rotation about the local tangent.  Thus we consider
pure crankshaft motion of the centerline about the axis of rotation.
Furthermore, the torsional stress $C{\bf r}_s\cdot{\mbox{\boldmath$\Omega$}}$
due to rotational drag of the held leg is ${\cal O}(\zeta_r\omega L)$,
much smaller than the torsional stress due to {\it translational} drag 
of the free leg, which for small $\omega$ is 
${\cal O}(\zeta_{\perp}\omega L^3)$.  As we shall verify below, the 
twist due to translation dominates that due to rotation even at large
$\omega$, as long as $L/a$ is sufficiently large.  Hence we set $\zeta_r=0$
for further simplicity.  Finally, we take the  
kink angle $\alpha$ to be clamped
at a value independent of $\omega$.
 
We desire the steady-state solution to~(\ref{omegas}--\ref{drag}) for 
${\bf r}_t={\mbox{\boldmath$\omega$}}\times{\bf r}$, 
with ${\mbox{\boldmath$\omega$}}=\omega{\bf\hat z}$; as our experiments confirm there is no oscillatory behavior as expected at low Reynolds number.  Fifteen boundary
conditions at the ends are required for the fifteen unknowns:
${\bf r}$, ${\bf r}_s$, ${\bf\hat e}_1$ (${\bf\hat e}_2={\bf r}_s
\times{\bf\hat e}_1$),
${\bf M}$, and ${\bf F}$.  Taking the $x$- and
$y$-axes to rotate about $z$ with rate $\omega$, we 
demand ${\bf r}(0)={\bf 0}$,
${\bf r}_s(0)={\bf\hat z}$, and ${\bf\hat e}_1(0)={\bf\hat x}$ at 
the held
end.  The free end experiences no force and moment: ${\bf F}(L)={\bf 0}$ and
${\bf M}(L)={\bf 0}$.  Finally, fifteen conditions must be enforced
at the position of the joint, $s=L_1$.  These conditions are the continuity
of position, force, and moment, and 
${\bf r}_s(L_1^+)={\bf r}_s(L_1^-)\cos\alpha
+{\bf\hat e}_1(L_1^-)\sin\alpha$, and
${\bf\hat e}_1(L_1^+)=-{\bf r}_s(L_1^-)\sin\alpha
+{\bf\hat e}_1(L_1^-)\cos\alpha$.
The twist is discontinuous across the kink, since the twisting
moment on one side can balance with a bending moment on the other; 
kinks trap torsional stress.

We measure length in units of $L$, time in units of 
the bending relaxation time $\zeta_\perp L^4/A$, 
force in units of $A/L^2$, and moment in units 
of $A/L$.  The shape of the rod
is then controlled by the dimensionless rotation 
rate $\chi=\zeta_{\perp}
\omega L^4/A$ ({\it cf.}~\cite{wrgo}) 
and the geometrical quantities $\alpha$ and $L_1/L$.
In our macroscopic experiments of Figs.~\ref{L}--\ref{V},  
the rod is a steel compression spring wrapped 
in Teflon$^{\tiny\rm TM}$ tape, immersed in glycerol, with
$\eta\approx20.0$ erg-sec/cm$^3$, $L=29$ cm, $a=0.16$ cm, 
$\alpha\approx87^\circ$, and
$A \approx C \approx 2.2 \times 10^5$ dyne-cm$^2$.  For our motor
speeds of $5$--$500$ rpm, $\chi$ ranges from $55$--$5500$.

For small $\chi$ the shape and stress can be obtained to linear order
in $\chi$.
For the linear calculation only, it is convenient to 
write (\ref{omegas}--\ref{drag}) in terms of ${\bf r}$ and 
set $\zeta_{\perp}=\zeta_{\parallel}$;
in fact, $\zeta_{\parallel}$ will not enter the linear calculation.
The constitutive relation, the moment balance 
equation~(\ref{mbalance}) (which now has ${\bf m}={\bf 0}$), 
and ${\bf r}_s\cdot{\bf r}_s=1$ imply
${\bf F}=-{\bf r}_{sss}+\Gamma\Omega{\bf r}_s\times{\bf r}_{ss}
+\Lambda{\bf r}_s$, where $\Omega={\bf r}_s\cdot{\mbox{\boldmath$\Omega$}}$, 
and $\Lambda$ is the unknown part of 
${\bf r}_s\cdot{\bf F}$.
Force balance is thus
\begin{equation}
\chi{\bf\hat z}\times{\bf r}=-{\bf r}_{ssss}
+\Gamma\Omega{\bf r}_s\times{\bf r}_{sss}
+(\Lambda{\bf r}_s)_s.
\label{rfbal}
\end{equation}
Since $\zeta_r=0$, the tangential
moment balance~(\ref{mbalance}) 
implies $\Omega_s=0$, {\it i.e.} constant twist in each leg.  

For simplicity, consider a right-angle kink, $\alpha=\pi/2$.
If $\chi=0$, then $\Omega=\Lambda=0$, and the rod is undeformed. To first 
order in $\chi$, the deformation of both legs of the rod is evidently
along the $y$-direction; therefore, 
\begin{equation}
{\bf r}(s)\approx\left\{\matrix{s{\bf\hat z}+y(s){\bf\hat y}&0\le s\le L_1\cr
L_1{\bf\hat z}+(s-L_1){\bf\hat x}+y(s){\bf\hat y}&L_1\le s\le1\cr}\right. .
\label{rperturb}
\end{equation}
Force balance implies that $\Lambda_s=0$ in both legs, $y_{ssss}=0$
for $0\le s\le L_1$, and $\chi(s-L_1)=-y_{ssss}$ for $L_1\le s\le1$.  
The boundary conditions at the ends of the rod are $y(0)=y_s(0)=0$,
and $y_{ss}(1)=y_{sss}(1)=\Omega(1)=\Lambda(1)=0$.  Therefore, 
$\Lambda=\Omega=0$ for $L_1<s\le 1$; note that the twist for $0\le s<L_1$ 
is {\it not} zero: the kink has trapped the twist.  
Continuity requires $y(L_1^-)=y(L_1^+)$; also,
\begin{equation}
\matrix{{\bf\hat e}_1^-
={\bf\hat x}+\Gamma\Omega L_1{\bf\hat y}& &{\bf\hat e}_1^+
=-{\bf\hat z}-y^+_s{\bf\hat y} \hfill\cr
{\bf\hat e}_2^-={\bf\hat y}-\Gamma\Omega L_1{\bf\hat x}&{\rm and}
&{\bf\hat e}_2^+={\bf\hat y}-y^+_s{\bf\hat x}+y^-_s{\bf\hat z} \cr
{\bf\hat e}_3^-={\bf\hat z}+y^-_s{\bf\hat y}\hfill& &{\bf\hat e}_3^+
={\bf\hat x}+y^+_s{\bf\hat y}\hfill\cr},
\label{frames}
\end{equation}
where ${\bf\hat e}_i^\pm={\bf\hat e}_i(L_1^\pm)$, {\it etc}.  
Using~(\ref{frames}),
the condition on the jump in the material frame becomes 
$\Gamma\Omega L_1=y^+_s$,
continuity of force becomes $y^-_{sss}=y^+_{sss}$ and $\Lambda^-=0$,
and continuity of moment becomes $y^-_{ss}=0$ and $\Gamma\Omega=y^+_{ss}$.
Thus we find a (dimensional) twist stress of 
$C\Omega=-\zeta_{\perp}\omega(L-L_1)^3/3$
in the held leg.  The (dimensional) shape of the rod is given by 
\begin{equation}
y(s)=-{\zeta_{\perp}\omega\over12A}s^2(3L_1-s)(L_1-L)^2,\quad s\le L_1,
\label{yless}
\end{equation}
and
\begin{eqnarray}
y(s)&=&{\zeta_{\perp}\omega\over A}\biggl[
\bigg({L_1^2L^3\over6}-{3L_1^3L^2\over4}+L_1^4L-{49L_1^5\over120}\bigg)
\nonumber\\
&+&L_1^2s\bigg({L^2\over4}-{L_1L\over2}+{5L_1^2\over24}\bigg)
+s^2L^2\bigg({L_1\over4}-{L\over6}\bigg)\nonumber\\
&+&s^3L\bigg({L\over12}-{L_1\over6}\bigg)+{s^4L_1\over24}
-{s^5\over120}\biggr]
,\quad s\ge L_1.
\label{ygreat}
\end{eqnarray}

To solve for the shape and stress when $\chi$ is not small, 
we must use numerical methods.  However,
we can obtain the $\chi$-dependence of the various quantities for $\chi\gg1$ 
using a simple argument.  Since the rod is mostly aligned
along the $z$-axis at high twirling rates (in both the folding and extending
cases), 
${\bf r}={\bf\hat z}s+{\bf r}_\perp$ with $|{\bf r}_\perp|\rightarrow0$ as 
$\chi\rightarrow0$.  Ignoring the terms involving $\Lambda$, and all 
numerical
prefactors, (\ref{rfbal}) reduces to
\begin{equation}
{\bf\hat z}\times{\bf r}_{\perp}=-{1\over\chi}{\bf r}_{\perp ssss}+
{\Omega\over\chi}
{\bf\hat z}\times{\bf r}_{\perp sss}+\cdots.
\label{rperp}
\end{equation}
The curvature and $|{\bf r}_\perp|$ are very small
except in a boundary layer near the kink at $s=L_1$; 
rescaling $s-L_1=({\bar s}-L_1)/\chi^{1/4}$, we find that the 
translational drag, bending, and twisting forces per 
length all balance for large $\chi$ if
$\Omega\propto\chi^{1/4}$.  On the other hand, we can
balance the moment $C\Omega$ against the approximate translational drag 
$ \int{\rm d} s (\zeta_{\perp} \omega|{\bf r}_{\perp}|) 
|{\bf r}_{\perp}|$ to 
find $|{\bf r}_{\perp}|\propto\chi^{-1/4}$.  This change in shape 
with $\chi$ implies that the {\it effective} rotational friction 
coefficient $\zeta_{r,{\rm eff}}\equiv C\Omega/\omega$ 
decreases as $\chi^{-3/4}$ for large $\chi$ .

Equations (\ref{omegas}--\ref{drag}) are
in standard form for the relaxation method~\cite{nr}.  The interval $0\le s
\le 1$ is replaced by a fine mesh, with two mesh points corresponding to 
$s=1/2^\pm$. There is no need to introduce the intermediate tension variable 
$\Lambda$, which enforces fixed length,
since (\ref{omegas}) and the boundary condition on ${\bf r}_s$ at $s=0$ 
imply 
${\bf r}_s\cdot{\bf r}_s=1$.  For simplicity, we choose $\Gamma=1$ and 
$\zeta_{\parallel}=\zeta_{\perp}/2$.

\begin{figure}
\centerline{\psfig{figure=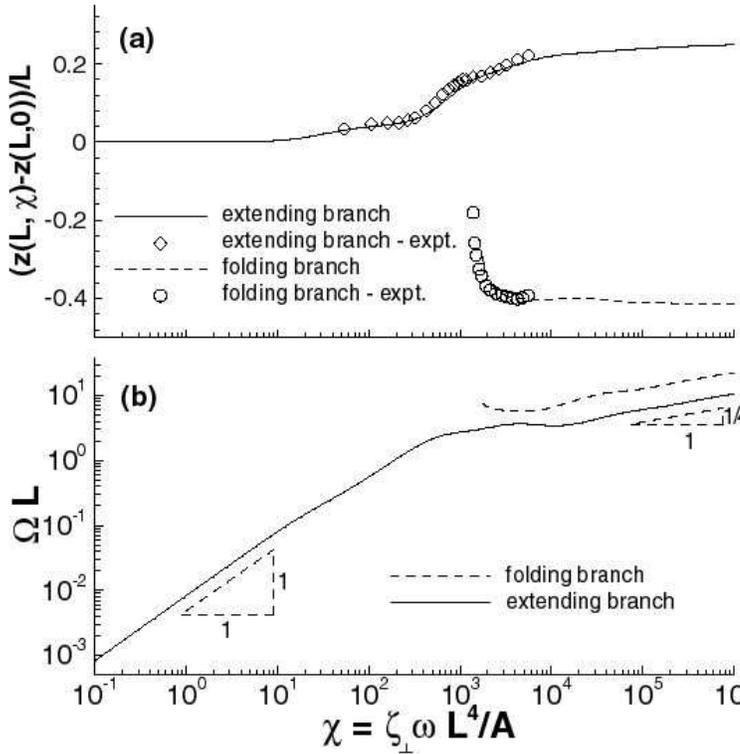,height=4.0in}}
\smallskip
\caption{(a) Theoretical predictions and experimental results
for relative $z$-displacement of rod free end ($s=L$) versus dimensionless
rotation rate $\chi$.
(b) Theoretical prediction for dimensionless twist in the held leg
versus $\chi$.  Note the agreement with the scaling
arguments for $\Omega$.  $\alpha\approx87^{\circ}$;
$L_1/L=2/3$. 
}
\label{graphs}
\end{figure}

Fig.~\ref{graphs} shows the total extension and torsional stress
as a function of twirling rate for a rod with $\alpha=87^\circ$ and 
$L_1/L=2/3$ (as in Figs.~\ref{L}--\ref{V}).  
As $\chi$ increases, the initially L-shaped rod distorts and extends.  
At sufficiently large $\chi$, a new branch of solutions appears which
corresponds to folding.  The situation is reminiscent of an imperfect
bifurcation.  To make the comparison with experiment, we have included
a net downward force per unit length due to gravity (the filament's linear 
density is $0.167$ g/cm and glycerol has a density around $\rho=1.3$ g/cm$^3$).
There is no fitting in Fig.~\ref{graphs}.
The agreement between theory and experiment is excellent.  We expect the 
slender-body hydrodynamics approximations to work well since the curvature is
gentle away from the small region near the kink (figs.~\ref{L}--\ref{V}), and
the two legs of the rod are separated by many rod radii even in the
folded state (inset to fig.~\ref{V}).  The  Reynolds number $Re\approx \omega r_{\rm max} a \rho/\eta$ is about 0.2 for $\omega$ around 500 rpm ($r_{\rm max}$ is the maximum transverse displacement of the rod, $\omega r_{\rm max}$ sets the velocity scale, and the rod {\it radius} $a$ sets the length scale of the disturbance flow).

\begin{figure}
\centerline{\psfig{figure=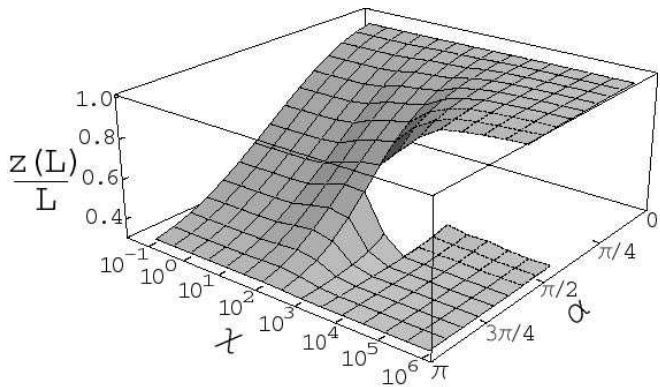,height=2.in}}
\smallskip
\caption{Phase diagram for twirling a rod with a kink at $L_1=2/3L$
(no gravity). For $\chi\protect\ageq1500$
and $\alpha$ near $\pi/2$,
the rod is bistable, as indicated by the dashed lines.
Here, $z=1$ is fully extended and $z=1/3$ is completely folded.
}
\label{phase}
\end{figure}

Fig.~\ref{graphs} also gives the dependence of the torsional 
stress on $\chi$, confirming our scaling arguments that $\Omega\propto\chi$
for $\chi\ll1$ and $\Omega\propto\chi^{1/4}$ for $\chi\gg1$.
We can now assess our neglect
of $\zeta_r$ at large $\chi$:  demanding 
$(\zeta_r\omega L)/(\int{\rm d} s \zeta_{\perp}\omega|{\bf r}_\perp|^2)\ll1$ 
amounts to requiring
$\chi^{3/4}(a/L)^2\ll1$, which is easily fulfilled for 
$L/a\approx10^3$.
Note that (within the slender-body hydrodynamics approximation) the twist
has the $\chi^{1/4}$ dependence for both the folding and extending
branches.

Fig.~\ref{phase} displays the ``phase-diagram'' for a 
twirling bent rod.  Rods with large kink angles
(V-shaped) fold up as $\chi$ increases from zero, whereas
rods with small kink angles extend.  For intermediate
kink angles, both states are possible at high $\chi$; there is no critical
angle dividing extending and folding.  This behavior is generic as long 
as $L_1/L$ is not too close to $0$ or $1$.

In conclusion, we have shown how kinks affect
the deformation and torsional stress of a twirling elastic rod in
a viscous fluid.  Despite the linearity of the elastic constitutive
relations (which derive from Hooke's law) and the equations for viscous
flow (Stokes equations), we find nonlinear dependences on rotation rate
at high rates due to the change in shape.  Bacterial flagella, DNA, 
and fibers of {\it B. subtilis} are sufficiently flexible that typical
rotation rates can cause the extending and folding studied here,
which could be revealed using micromanipulation.  An important extension
of our work which would be relevant to DNA would be to carry out Brownian
dynamics simulations of a rotating flexible rod with many kinks, as 
envisioned in ref.~\cite{pcn}.

We thank P. Nelson for posing the question that led to this work,
and R. Goldstein, T. Peacock, H. Stone, C. Wiggins, 
and C. Wolgemuth  for discussions.  We thank H. Stone for partial support through
the Harvard MRSEC and the Army Research Office Grant DAA655-97-1-014.

\end{multicols}

\end{document}